\documentclass[12pt]{iopart}

\usepackage{graphicx}  
\newcommand{\snn}{\sqrt{s_{NN}}}

\newcommand{\pbar}{\overline{p}}

\newcommand{\nch}{N_{ch}}
\newcommand{\np}{N_{part}}

\newcommand{\dndeta}{d\nch/d\eta}

\begin{document}

\title{The Landscape of Particle Production: Results from PHOBOS}

\author{Peter Steinberg$^1$ for the PHOBOS Collaboration}
%
%
\noindent
B B Back$^2$,
M D Baker$^1$,
M Ballintijn$^3$,
D S Barton$^1$,
B Becker$^1$,
R R Betts$^4$,
A A Bickley$^5$,
R Bindel$^5$,
A Budzanowski$^6$,
W Busza$^3$,
A Carroll$^1$,
M P Decowski$^3$,
E Garc\'{\i}a$^4$,
T Gburek$^6$,
N George$^{1,2}$,
K Gulbrandsen$^3$,
S Gushue$^1$,
C Halliwell$^4$,
J Hamblen$^7$,
A S Harrington$^7$,
G A Heintzelman$^1$,
C Henderson$^3$,
D J Hofman$^4$,
R S Hollis$^4$,
R Ho\l y\'{n}ski$^6$,
B Holzman$^1$,
A Iordanova$^4$,
E Johnson$^7$,
J L Kane$^3$,
J Katzy$^{3,4}$,
N Khan$^7$,
W Kucewicz$^4$,
P Kulinich$^3$,
C M Kuo$^8$,
J W Lee$^3$,
W T Lin$^8$,
S Manly$^7$,
D McLeod$^4$,
A C Mignerey$^5$,
R Nouicer$^{1,4}$,
A Olszewski$^6$,
R Pak$^1$,
I C Park$^7$,
H Pernegger$^3$,
C Reed$^3$,
L P Remsberg$^1$,
M Reuter$^4$,
C Roland$^3$,
G Roland$^3$,
L Rosenberg$^3$,
J Sagerer$^4$,
P Sarin$^3$,
P Sawicki$^6$,
I Sedykh$^1$,
W Skulski$^7$,
C E Smith$^4$,
G S F Stephans$^3$,
A Sukhanov$^1$,
J -L Tang$^8$,
M B Tonjes$^5$,
A Trzupek$^6$,
C Vale$^3$,
G J van~Nieuwenhuizen$^3$,
R Verdier$^3$,
G I Veres$^3$,
F L H Wolfs$^7$,
B Wosiek$^6$,
K Wo\'{z}niak$^6$,
A H Wuosmaa$^2$,
B Wys\l ouch$^3$ and
J Zhang$^3$\\
%
%
%
%
\address{
$^1$~Brookhaven National Laboratory, Upton, NY 11973-5000, USA\\
$^2$~Argonne National Laboratory, Argonne, IL 60439-4843, USA\\
$^3$~Massachusetts Institute of Technology, Cambridge, MA 02139-4307, USA\\
$^4$~University of Illinois at Chicago, Chicago, IL 60607-7059, USA\\
$^5$~University of Maryland, College Park, MD 20742, USA\\
$^6$~Institute of Nuclear Physics, Krak\'{o}w, Poland\\
$^7$~University of Rochester, Rochester, NY 14627, USA\\
$^8$~National Central University, Chung-Li, Taiwan\\
}

\begin{abstract}
Recent results from the PHOBOS experiment at RHIC are presented,
both from Au+Au collisions from the 2001 run
and p+p and d+Au collisions from 2003.  
The centrality dependence of the total charged particle
multiplicity in p+p and d+Au show features, such as $\np$-scaling
and limiting fragmentation, similar to $p+A$ collisions at lower energies. 
Multiparticle physics in Au+Au is found to be local in (pseudo)rapidity,
both when observed by HBT correlations and by forward-backward
pseudorapidity correlations.
The shape of elliptic flow in Au+Au, measured over the full range of pseudorapidity,
appears to have a very weak centrality dependence.
Identified particle ratios in d+Au reactions 
show little difference between the shape of proton
and anti-proton spectra, while the absolute yields show an approximate
$m_T$ scaling.
Finally, results on $R_{dAu}$ as a function of pseudorapidity,
show that this ratio decreases monotonically with $\eta$, even
between $0.2<\eta<1.4$.
\end{abstract}




\begin{figure}[t]
\begin{center}
\includegraphics[height=7cm]{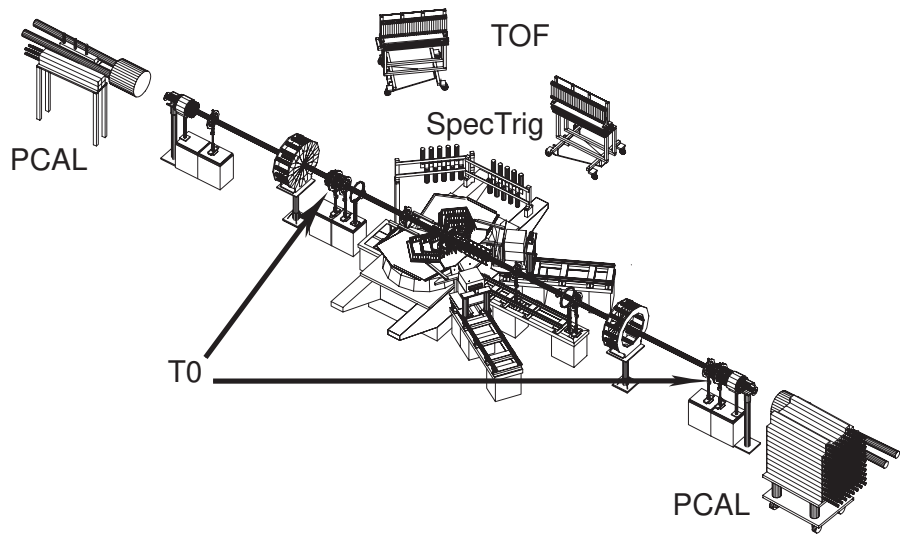}
\end{center}
\caption{
The PHOBOS detector configuration for the 2003
d+Au run. 
}
\end{figure}

\section{Introduction}
The PHOBOS detector at RHIC was designed to offer an overview
of particle production in the collisions of a variety of
systems (p+p, d+Au, Au+Au) over a wide range of control
parameters: beam energy, pseudorapidity, and centrality.  

The PHOBOS apparatus
was upgraded substantially for the 2003 d+Au RHIC run.  
Two calorimeters (PCAL) were added 
to provide information on spectator protons
from the d and Au projectiles. 
Another major upgrade involved the PHOBOS particle-identification
system.
The two time-of-flight (TOF) walls were moved back from their design
position to improve the mass resolution.  
Time resolution was improved by the addition of two sets of 10
Cerenkov timing start counters (T0).
Most importantly, two small hodoscopes were added just behind
the PHOBOS spectrometer to provide the means to trigger on
high-momentum particles which also hit the TOF walls.  
This spectrometer trigger (SpecTrig) was installed to enrich
the sample of identified particles at high transverse momentum
($p_T = 3-4$ GeV/c).
 
\section{Charged-Particle Multiplicity}

\begin{figure}[t]
\begin{center}
\begin{minipage}{70mm}
\includegraphics[width=7cm]{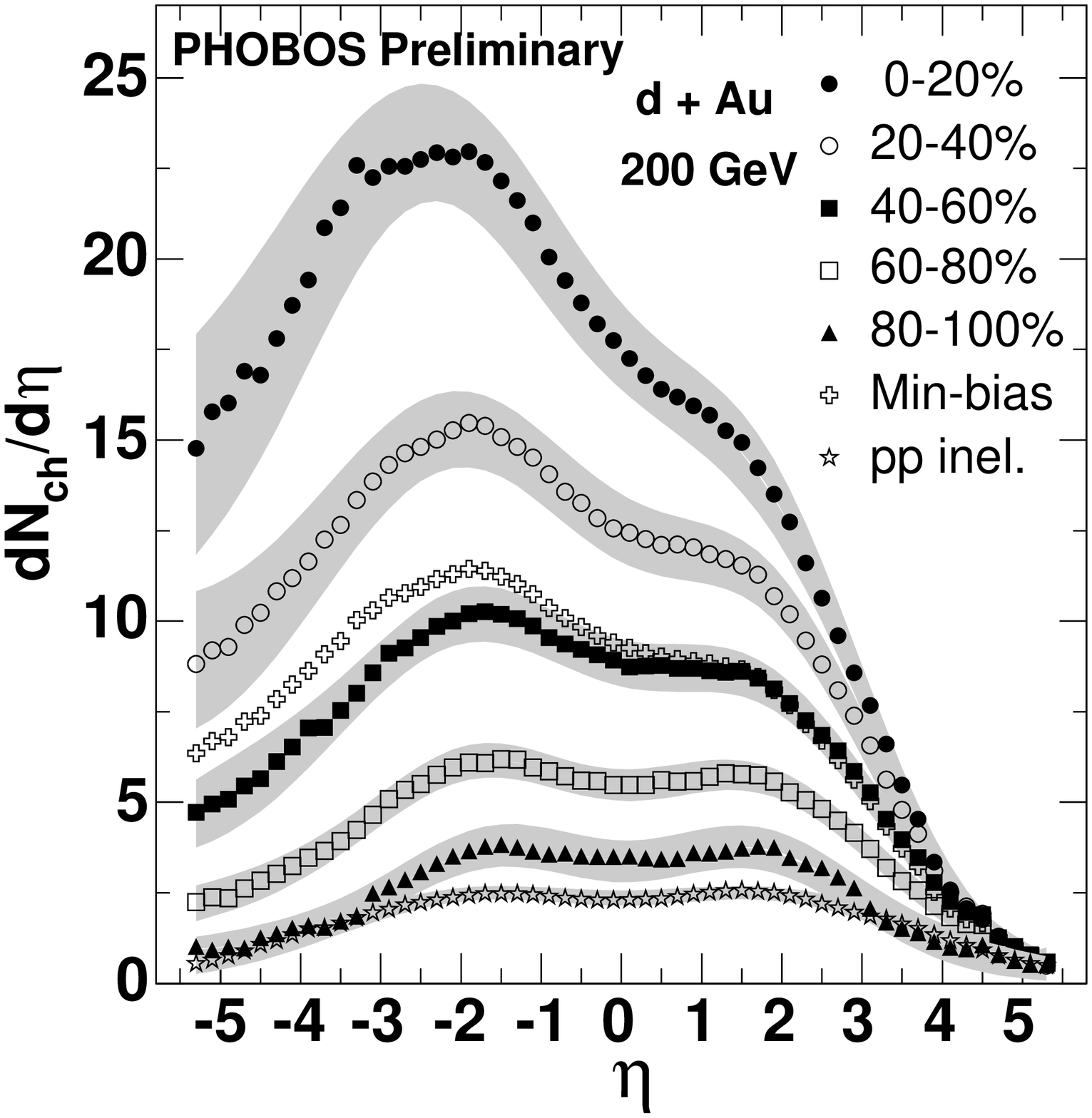}
\caption{$\dndeta$ for centrality bins in p+p and d+Au
collisions.  Bands are 90\% C.L.  systematic errors.
Error bars on min-bias data are suppressed for clarity.
\label{daudndeta}}
\end{minipage}
\hspace{\fill}
\begin{minipage}{80mm}
\includegraphics[width=8cm]{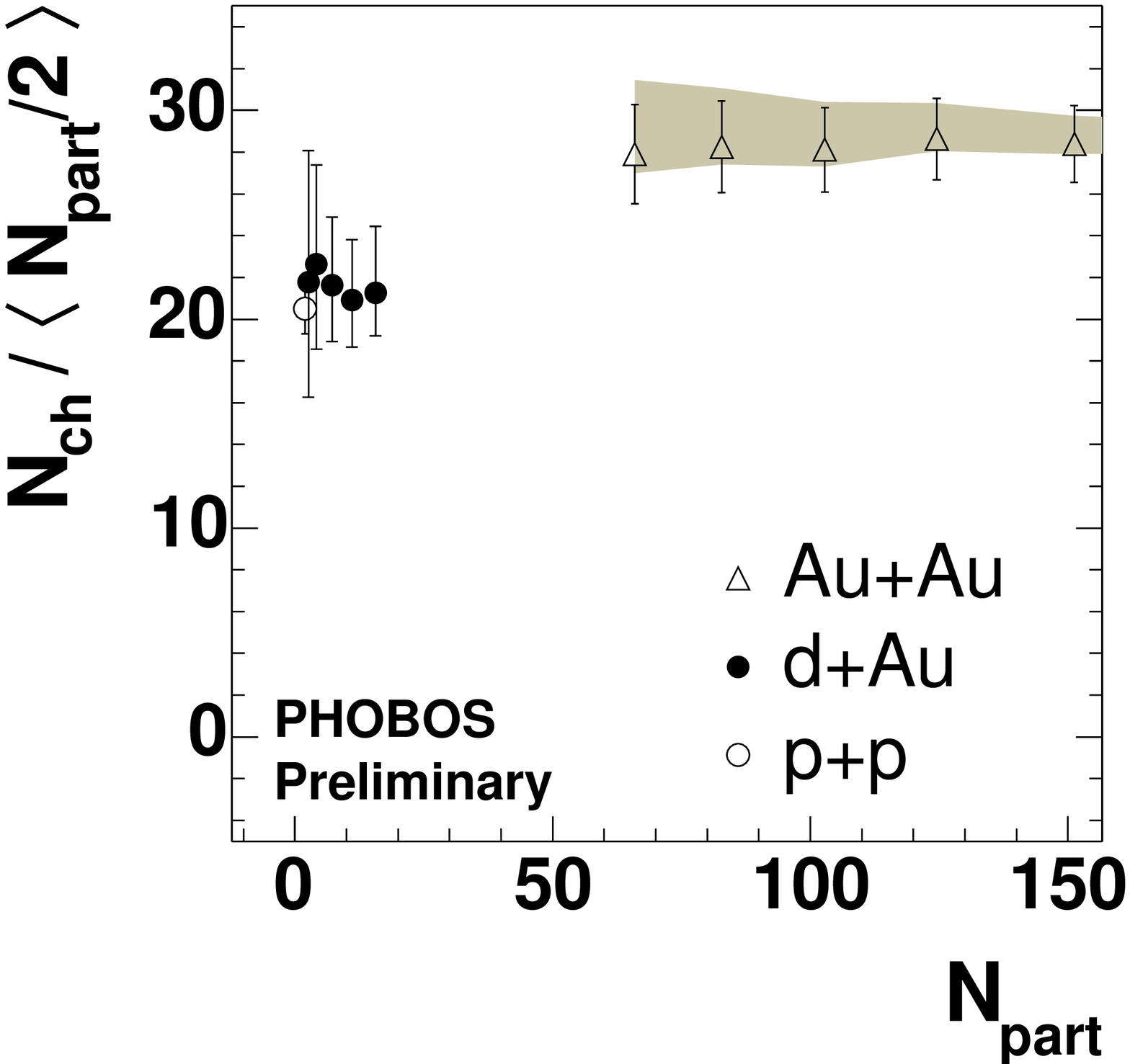}
\caption{
$N^{d+Au}_{ch}$ (integrated over $4\pi$) divided
by $\langle N_{part}/2 \rangle$ vs. $\np$.  Error bars
are 90\% C.L. while the grey band expresses the systematic
error on $\np$ in Au+Au.
\label{dauntot}
}
\end{minipage}
\end{center}
\end{figure}


In 2003, PHOBOS took data at $\snn= 200$ GeV for both 
p+p and d+Au collisions.  Minimum-bias d+Au data has already
been released \cite{dau_mult}.  As with the d+Au data, the p+p
data were corrected for trigger bias in an attempt to estimate the
pseudorapidity distribution for the total inelastic cross section.
Using the signals in the PHOBOS ring detectors ($3<|\eta|<5.4$),
the d+Au data has been divided into centrality bins, which 
are shown in Fig. \ref{daudndeta}.
As with Au+Au collisions, both limiting fragmentation relative to
lower-energy proton-nucleus data \cite{busza,elias}, 
and participant scaling of the
total multiplicity \cite{phobos-univ} are observed, 
the latter being shown in Fig. \ref{dauntot}.  
These results are somewhat surprising
since different dynamical effects could be expected to dominate
in the deuteron and gold hemispheres, such as energy stopping and
secondary cascading.  The robustness of the $\np$ scaling perhaps suggests
some sort of long-range correlation among the produced particles.
The different constant of proportionality between the p+p/d+Au
data and the Au+Au data also suggests a fundamental difference
between the two, once moderate centralities are reached in Au+Au.
These results, along with the detailed energy dependence of
pseudorapidity distributions in p+A and d+A, 
are discussed in more detail in Ref. \cite{nouicer}.

\section{Directed and Elliptic Flow in Au+Au}

\begin{figure}[t]
\begin{center}
\begin{minipage}{80mm}
\includegraphics[height=6cm]{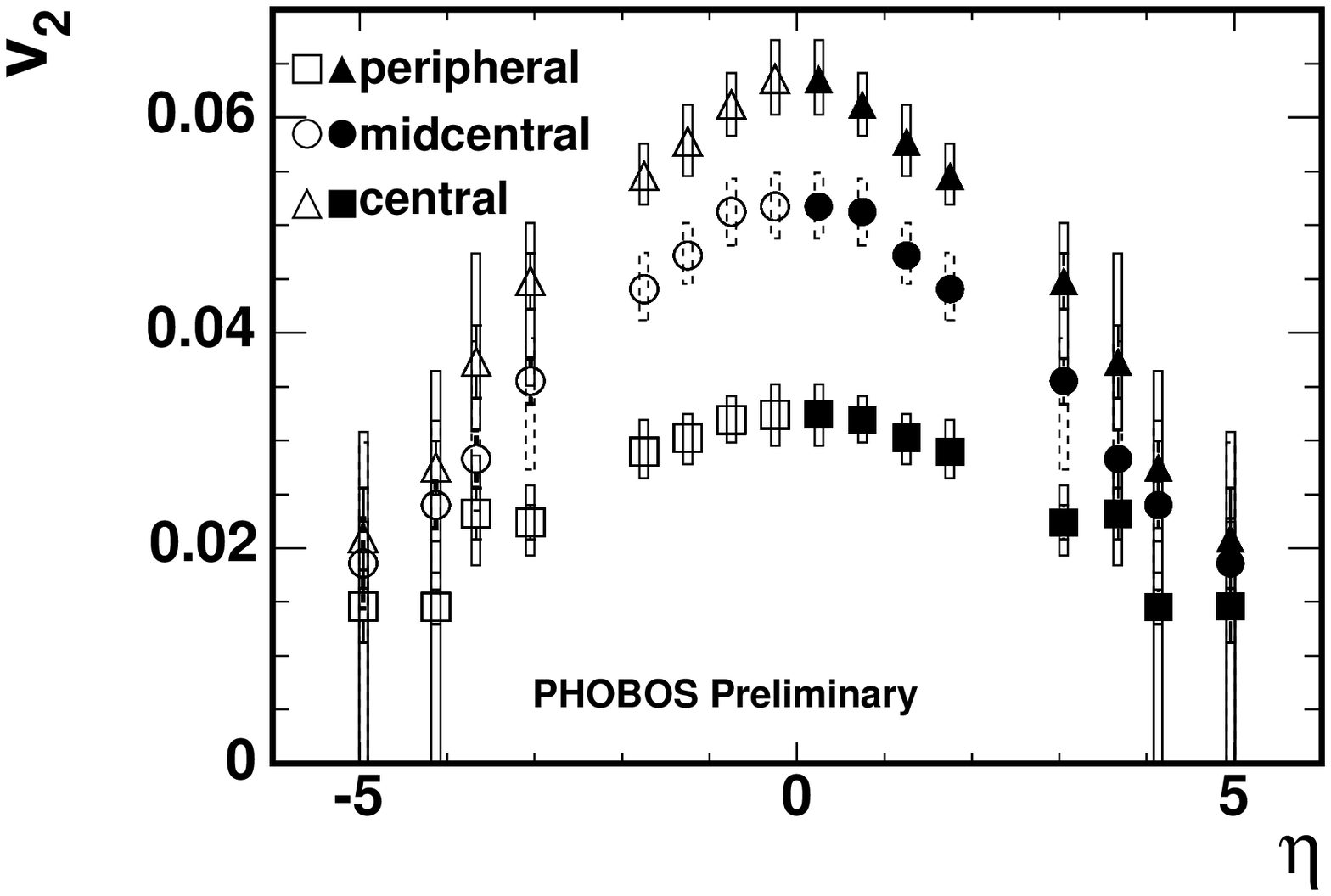}
\caption{$v_2$ for 3 centrality bins in 200 GeV Au+Au collisions.
Statistical errors are shown as bars.
90\% C.L. systematic errors are indicated by boxes.
\label{v2cent}
}
\end{minipage}
\hspace{\fill}
\begin{minipage}{60mm}
\includegraphics[height=6cm]{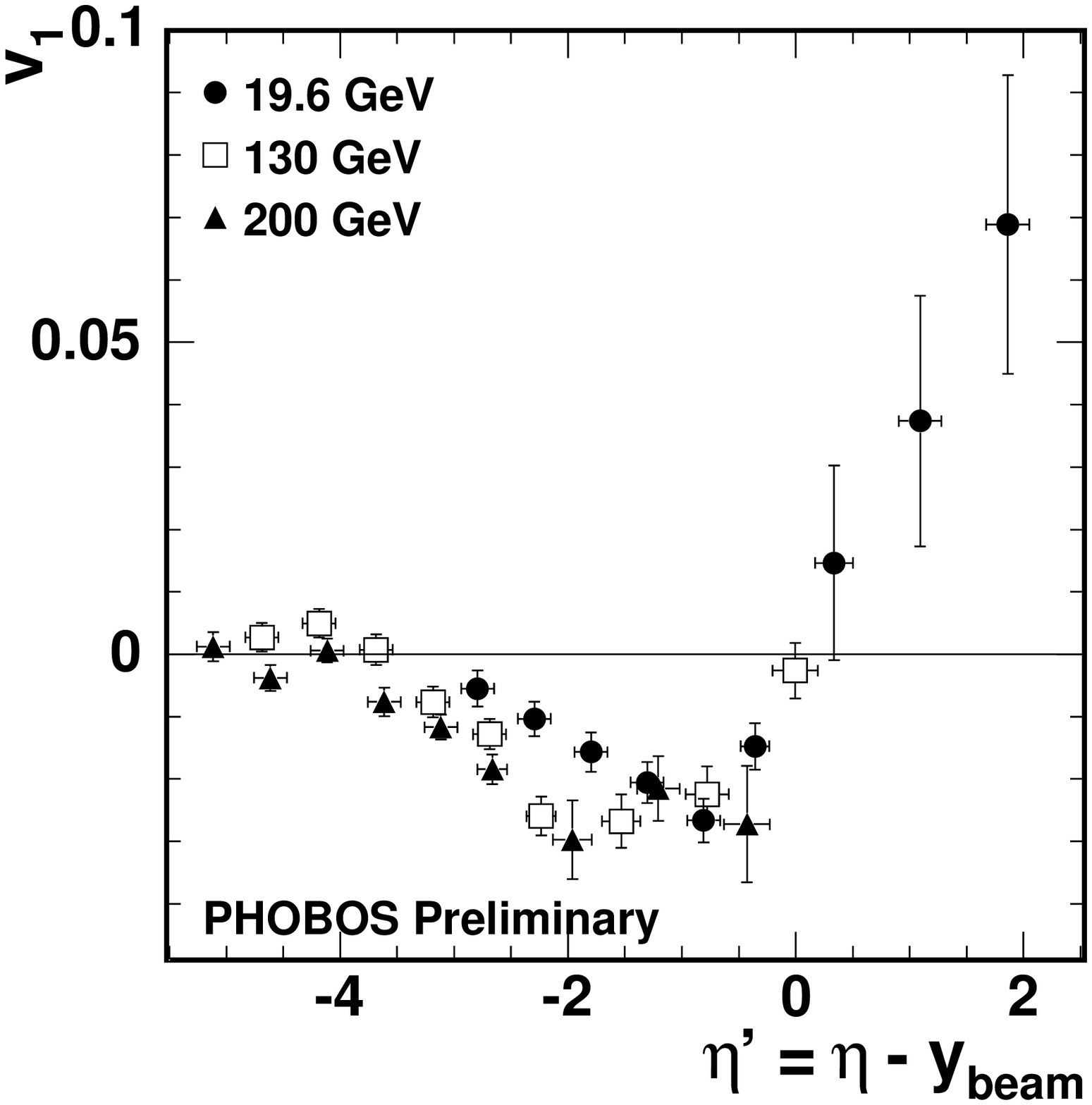}
\caption{
$v_1$ as a function of $\eta^{\prime}$ for Au+Au collisions at three energies.
Statistical errors are shown as bars.
90\% C.L. systematic errors are suppressed in this figure.
\label{v1eprime}
}
\end{minipage}
\end{center}
\end{figure}

PHOBOS results on $v_2$ as a function of pseudorapidity
have been shown previously for data integrated over all
available centralities \cite{phobos-flow-130,phobos-flow-qm02}.  
Preliminary results have extended this
measurement by a division of the data into three centrality
bins, shown in Fig. \ref{v2cent}.  These results show that 
the shape is manifestly not boost invariant over all $\eta$
in peripheral collisions (although one must not ignore kinematic
effects near midrapidity, as pointed out by Kolb \cite{kolb}).
Moreover, the shape does not change dramatically as a function of 
centrality, apart from an overall scale factor.

New at QM2004 are results on $v_1$, the first harmonic of the
azimuthal angular distribution, as a function of pseudorapidity and
beam energy.
A simple picture, which considers the effect of the net baryons,
would suggest that $v_1$ should develop opposite signs for
baryons and mesons, with the baryons defined to have a positive
directed flow signal near projectile rapidities.
Our results on $v_1$ show a striking energy dependence, which
leads to a large range in $\eta$ with $v_1\sim 0$ for the 
higher energies and then
a small but systematic negative $v_1$ that is approximately independent of
energy when viewed relative to beam rapidity, shown in Fig. \ref{v1eprime}.
This may be apparently another manifestation
of ``limiting fragmentation'', already seen in the pseudorapidity
distributions of charged particles \cite{phobos-limfrag}.
Studies such as these should
be useful in understanding the global dynamics of Au+Au collisions.
These results, along with comparisons to other experimental data,
are explained in more detail in Ref. \cite{belt}.

\section{Forward Multiparticle Physics}

\begin{figure}[t]
\begin{center}
\begin{minipage}{7cm}
\includegraphics[width=7cm]{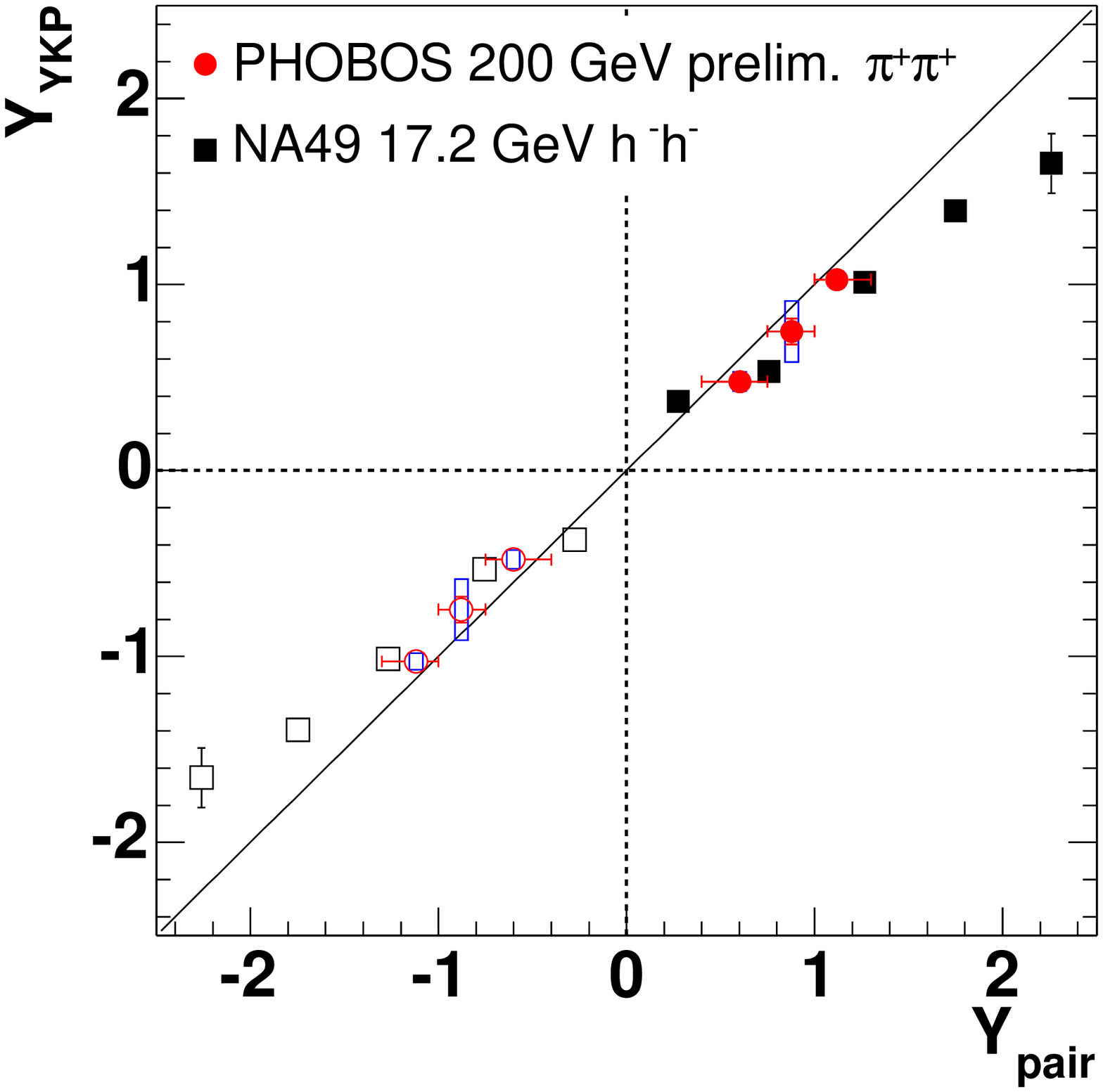}
\caption{Correlation of the YKP source rapidity with the
average rapidity of the pion pair.
\label{yykp}}
\end{minipage}
\hspace{\fill}
\begin{minipage}{7cm}
\includegraphics[height=7cm]{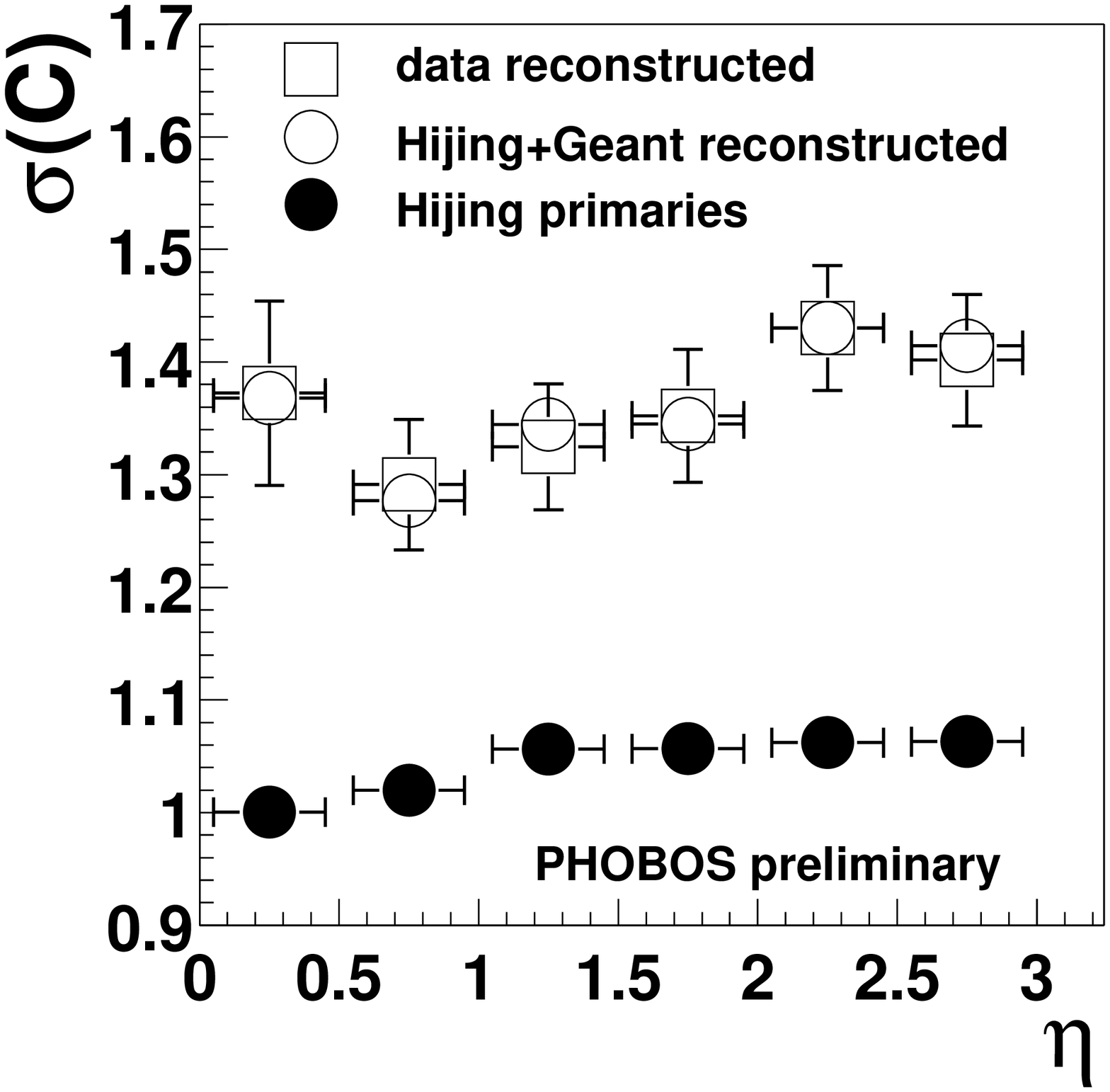}
\caption{
Forward backward correlation measure $\sigma(C)$ vs. $\eta$
for a bin width of $\Delta\eta = 0.5$.  
\label{sigmac}
}
\end{minipage}
\end{center}
\end{figure}

Another approach to understanding the role of  
correlations in rapidity space
is the use of multi-particle correlation measures like
two-particle HBT correlations and forward-backward correlations.  

PHOBOS has performed the first measurement of the rapidity dependence of HBT
correlations at RHIC.  
Using the YKP (Yano-Koonin-Podgoretskii) parametrization, we have also
been able to correlate the rapidity of the local source as a function
of the average rapidities of the two particles.
Shown in Fig. \ref{yykp}, one sees that these two rapidities are
tightly correlated, suggesting that particles produced at a certain
rapidity were produced by a source moving collectively at the
same rapidity.   This result is similar that found by the 
NA49 experiment at the SPS.
These results are explained in more detail in Ref. \cite{holzman}.

Another indication that multiparticle correlations reflect 
physics local in rapidity 
comes from the measurement of forward-backward correlations.
One would expect a strong correlation between particles produced
in rapidity bins in opposite hemispheres, simply from the fact that particle
production is controlled by the number of participants.
The quantitative
analysis of relative fluctuations can be used to assess the level
of short range correlations, which should lead to a non-binomial
distribution of multiplicities between two identical bins in the forward ($P$)
and backward ($N$) directions.  
By analyzing
the variance of a variable $C = (P-N)/\sqrt{P+N}$, $\sigma(C)$, 
as a function of the total multiplicity $P+N$, 
compared to a detailed Monte Carlo simulation of our apparatus, we can study 
non-binomial contributions to the fluctuations.
In Fig. \ref{sigmac}, we show data on $\sigma(C)$ for a bin width
$\Delta \eta = 0.5$ vs. $\eta$ as open squares.
Results from HIJING events are shown for
primary charged particles (closed circles) and full simulations (open
circles). 
The deviations of $\sigma(C)$ for the primary charged particles from unity
indicate that short-range correlations exist with a mild
pseudorapidity dependence for $\eta>1$.  The agreement of the data
with the fully-simulated HIJING suggests similar correlations are
present in our data set.
These results are shown for a different values of $\Delta\eta$ 
and discussed in detail in Ref. \cite{wozniak}.

\begin{figure}[t]
\begin{center}
\begin{minipage}{90mm}
\includegraphics[width=8cm]{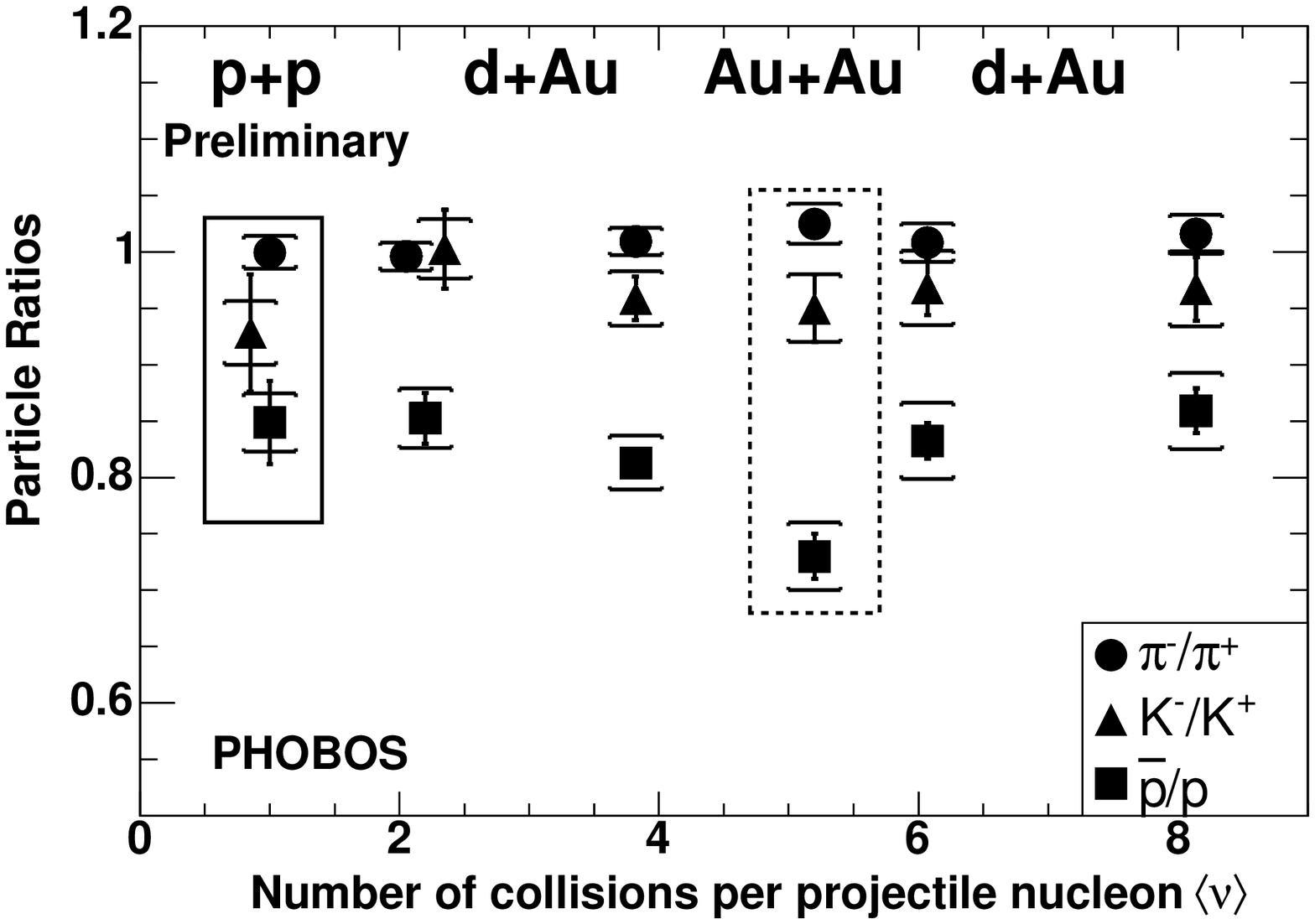}
\caption{Identified particle ratios in 200 GeV
p+p, d+Au, and Au+Au collisions as a function of $\nu$.
Statistical errors are shown as bars.
90\% C.L. systematic errors are indicated by brackets.
\label{ratiosvsnu}}
\end{minipage}
\hspace{\fill}
\begin{minipage}{60mm}
\includegraphics[width=6cm]{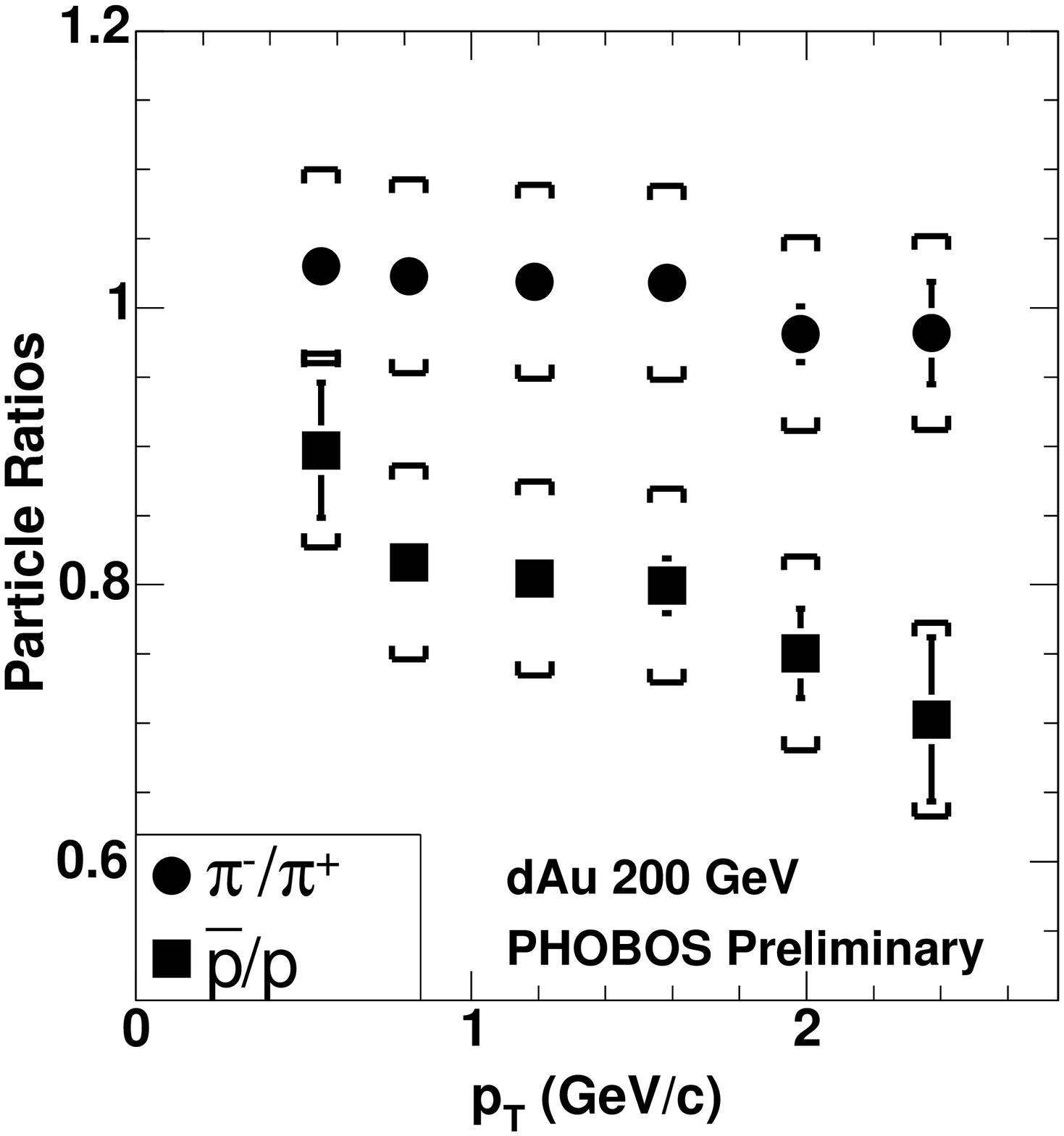}
\caption{
$\pi^-/\pi^+$ and $\pbar/p$ ratios vs. $p_T$
in 200 GeV d+Au collisions. 
Statistical errors are shown as bars.
90\% C.L. systematic errors are indicated by brackets.
\label{ratiosvspt}
}
\end{minipage}
\end{center}
\end{figure}

\section{Identified Particle Spectra}

One of the major upgrades to the PHOBOS detector in 2003 was
the rearrangement of the TOF walls and the addition of a trigger 
to increase our statistics of events with a high-momentum
track in the spectrometer.
With the enriched data set, we were able to make several new
measurements of identified particles at high $p_T$ in d+Au collisions.
These results are explained in more detail in Ref. \cite{veres}.

Ratios of identified particles and anti-particles are a tool
to study baryon transport.  Various models predict a substantial
transport of the initial baryon number of the deuteron all
the way to mid-rapidity, leading to a progressive decrease in
the $\overline{p}/p$ ratio as a function of the number of
collisions of each deuteron participant (characterized by
$\nu$) \cite{dau_pbarp}.  Our results for this ratio using $dE/dx$ PID, shown
in Fig. \ref{ratiosvsnu}, show
surprisingly little centrality dependence for particle momenta below
1 GeV/c and essentially the same result is found in p+p and d+Au.  
Using the TOF to extend our particle identification from
1.75 GeV/c to 3 GeV/c, this ratio shows a similarly weak centrality dependence.
Moreover, this ratio is also weakly dependent on $p_T$, 
as seen in Fig. \ref{ratiosvspt}.  This suggests that proton
and anti-proton spectra are similar, except for a difference in overall yield.

\begin{figure}[t]
\begin{center}
\includegraphics[height=6cm]{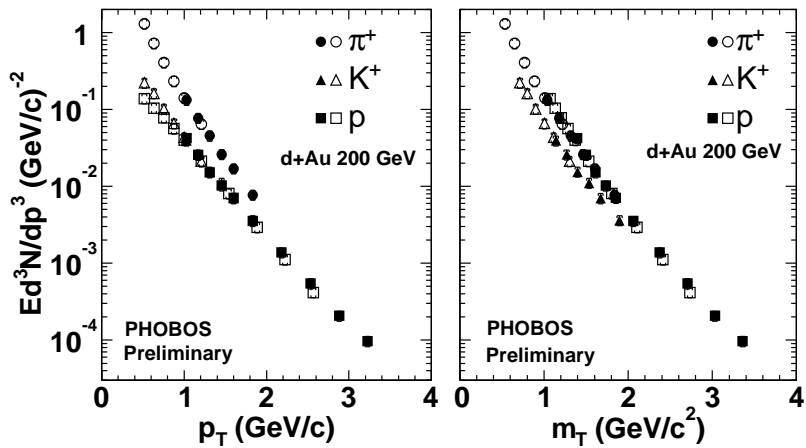}
\end{center}
\caption{Absolute yields of positively-charged
identified particles in 200 GeV
d+Au collisions, shown as a function of $p_T$ (left)
and $m_T$ (right).  The open and closed symbols represent
two different bending directions.
The data are not corrected for contributions
from weak decays.  
Statistical errors are shown as bars.
90\% C.L. systematic errors are indicated by brackets.
\label{pt_mt_pos}}
\end{figure}

The increased statistics in the spectrometer has enabled PHOBOS
to measure spectra of identified pions, kaons, and
protons of both charge signs up to 3.5 GeV.  The presence of
a ``Cronin effect'' in d+Au collisions at RHIC energies
has raised the question of whether any particular species is
responsible for this effect.
We have found that 
the relative contributions to the inclusive spectra vary
substantially with $p_T$, especially the growing fraction of
protons and anti-protons, as shown by the invariant yields
plotted vs. $p_T$ in Fig. \ref{pt_mt_pos}.  
However, when the yields are shown as a function of $m_T$, the
relative contributions appear to have little $m_T$-dependence.  
In other words, an apparent $m_T$ scaling is evident in the $d+Au$ data.
This is also seen in $p+p$ data, but is dramatically
violated at low momentum in Au+Au \cite{stopping}.

\section{Inclusive Spectra vs. Pseudorapidity}

\begin{figure}[t]
\begin{center}
\includegraphics[height=7cm]{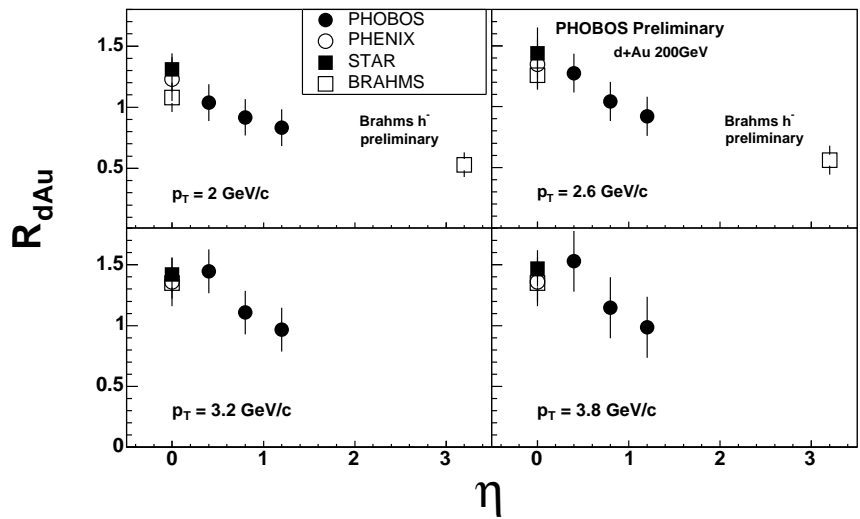}
\caption{
$R_{dAu}$ at fixed values of $p_T$ shown as a function
of $\eta$, compared with results from other RHIC experiments.
Error bars are statistical only.
\label{rdauvseta}
}
\end{center}
\end{figure}

One of the more striking results from the 2003 data 
is the observation of a significant
$\eta$ dependence in the modification of $d+Au$ spectra 
relative to $p+p$ data \cite{brahms,dau_cronin}.
This has been interpreted as
an indication of the onset of parton saturation \cite{cgc}
in the forward region.
To address this topic, PHOBOS has repeated its measurement of 
$R_{dAu}$, originally integrated over our full acceptance
($0.2 < \eta < 1.4$) \cite{dau_cronin}, but now in three bins ($0.2<\eta<0.6$, 
$0.6 <\eta<1.0$,
and 
$1.0<\eta<1.4$).  
Even within the acceptance of the PHOBOS spectrometer, $\Delta\eta=1.2$ units,
a dramatic drop in this ratio is seen as a function of $\eta$.  
When compared with 
other results from RHIC in Fig. \ref{rdauvseta},
it is found that the suppression seen at
forward rapidities is a continuation of a trend started
already at midrapidity.  Measurements in the backward hemisphere
by PHENIX show a clear enhancement of pion production with 
$1< p_T <3$ GeV/c \cite{phenix}, suggesting that this is not an effect that
``turns on'' in the forward region, but rather part of a difference
between $p+p$ and $d+Au$ data that extends over a large range in $\eta$.

\section{Conclusions}
Recent p+p, d+Au, and Au+Au results at 200 GeV
from the PHOBOS experiment at RHIC have been presented.
The systematic measurements shown here should serve as
a useful baseline for understanding the differences between
heavy ion collisions and more elementary systems.

\ack
%
%
%
%
This work was partially supported by U.S. DOE grants 
DE-AC02-98CH10886,
DE-FG02-93ER40802, 
DE-FC02-94ER40818,  
DE-FG02-94ER40865, 
DE-FG02-99ER41099, and
W-31-109-ENG-38, by U.S. 
NSF grants 9603486, 
0072204,            
and 0245011,        
by Polish KBN grant 2-P03B-10323, and
by NSC of Taiwan under contract NSC 89-2112-M-008-024.

\end{document}